\documentclass[fleqn,10pt]{wlscirep}
\usepackage[utf8]{inputenc}
\usepackage[T1]{fontenc}
\usepackage{epstopdf}
\usepackage{xcolor}
\usepackage{amsmath,amssymb,graphicx}
\usepackage{blindtext}

\title{Surpassing the Classical Limit in Magic Square Game with Distant Quantum Dots Coupled to Optical Cavities}

\author[1,*]{Sinan Bugu}
\author[2,3]{Fatih Ozaydin}
\author[1]{Tetsuo Kodera}
\affil[1]{Department of Electrical and Electronic Engineering, Tokyo Institute of Technology, 2-12-1 Ookayama,
Meguro-ku, Tokyo 152-8552, Japan}
\affil[2]{Institute for International Strategy, Tokyo International University, 1-13-1 Matoba-kita, Kawagoe, Saitama, 350-1197, Japan}
\affil[3]{Department of Information Technologies, Isik University, Sile, Istanbul, 34980, Turkey}
\affil[*]{bugu.s.aa@m.titech.ac.jp}

\begin{abstract}
The emergence of quantum technologies is heating up the debate on quantum supremacy, usually focusing on the feasibility of looking good on paper algorithms in realistic settings, due to the vulnerability of quantum systems to myriad sources of noise.
In this vein, an interesting example of quantum pseudo-telepathy games that quantum mechanical resources can theoretically outperform classical resources is the Magic Square game (MSG), in which two players play against a referee. 
Due to noise, however, the unit winning probability of the players can drop well below the classical limit.
Here, we propose a timely and unprecedented  experimental setup for quantum computation with quantum dots inside optical cavities, along with ancillary photons for realizing interactions between distant dots to implement the MSG. 
Considering various physical imperfections of our setup, we first show that the MSG can be implemented with the current technology, outperforming the classical resources under realistic conditions.
Next, we show that our work gives rise to a new version of the game. 
That is, if the referee has information on the physical realization and strategy of the players, he can bias the game through filtered randomness, and increase his winning probability. 
We believe our work contributes to not only quantum game theory, but also quantum computing with quantum dots.

\end{abstract}
\begin{document}

\flushbottom
\maketitle

\thispagestyle{empty}

		\section*{Introduction}

\noindent Quantum mechanical resources can enable some tasks such as superdense coding and teleporting an unknown state \cite{nielsen2001quantum}  which are impossible to realize with classical resources. Many approaches to optimizing quantum resources for efficient quantum computation and quantum communication such as gate-model, quantum channel capacity, optimizing quantum memory, and algorithms have been studied \cite{gyongyosi2019survey,gyongyosi2018survey,gyongyosi2020optimizing,gyongyosi2020circuit,gyongyosi2020quantumstate,laszlo2019dense,gyongyosi2019quantum,gyongyosi2019training,farhi2020quantum,farhi2020quantumapproximate,lloyd2018quantum}.
On the other hand, speeding up classically possible computational tasks which are beyond the ability of any classical computer such as unsorted database search and factorization \cite{nielsen2001quantum} and some other devoted efforts\cite{harrow2017quantum,neill2018blueprint,bremner2017achieving}  in achieving supremacy have been attracting an intense attention.
One of the most groundbreaking advances in quantum technologies is the recent claim of Google that they have achieved quantum supremacy \cite{arute2019quantum}.

Surpassing the classically achievable limit in various tasks is also in the center of attraction. For example in quantum metrology, surpassing the classical shot noise limit has been studied extensively under various scenarios taking into account the standard decoherence channels and thermal noise \cite{PezzeSmerzi2009PRL,Ma2011PRA,Erol2014SciRep,Ozaydin2014PLA,AliAoP,Ozaydin2015SciRep1,Ozaydin2020OQEL}.
Quantum resources also enable advantages in thermodynamics  \cite{scully2003extracting,turkpencce2016quantum,tuncer2019work,dag2019temperature}.
Quantum games -where ``everyone wins'' \cite{EveryoneWins}, provide an interesting playground for investigating the advantages of utilizing various quantum weirdness over classical resources. 
Among quantum pseudo-telepathy games where quantum mechanical resources can theoretically outperform classical resources, a widely studied one is the so-called Magic Square game (MSG), in which two players, say Alice and Bob, play against a referee. In the MSG, players are allowed to communicate, share any resources and agree on any strategy, only until the game starts.
The game is played on a $3 x 3$ square matrix with binary entries.
Once the game starts, referee gives numbers $a$ and $b$ to Alice and Bob, respectively, where $a, b \in \{1, 2, 3\}$.
Alice fills row $a$ and Bob fills column $b$, i.e. each tell referee the numbers to fill.
They win if the sum of numbers in row $a$ (column $b$) is even (odd) and the  intersecting element is the same. 
Otherwise, they lose, i.e. the referee wins.
Let us illustrate one of the nine possible instances that referee gives Alice $a=2$, and Bob $b=3$. 
They will win if they can fill the row and column as $\{0,0,0\},\{0,0,1\}$, respectively, or $\{0,1,1\},\{0,1,0\}$, for example, resulting in two possible winning instances (i) and (ii) given in Table ~\ref{table:ExampleWin}.

\begin{table}[t!]
	\centering
	\begin{tabular}{|c|ccc|l}
		\cline{1-4}
		i &                       	& Bob                    			&   						&  \\ \cline{1-4}
		& \multicolumn{1}{c|}{}  	& \multicolumn{1}{c|}{}  			&  \ \  \ \ 0  \ \ 	 \ \ 			&  \\ \cline{2-4}
		Alice 						& \multicolumn{1}{c|}{ \ \ 0 \ \ } 	& \multicolumn{1}{c|}{0} 	& 0 	&  \\ \cline{2-4}
		& \multicolumn{1}{c|}{}  	& \multicolumn{1}{c|}{}  			& 1 						&  \\ \cline{1-4}
	\end{tabular}\!\!\!\!, \ \ 
	\begin{tabular}{|c|ccc|l}
	\cline{1-4}
	ii &                       	& Bob                    			&   						&  \\ \cline{1-4}
	& \multicolumn{1}{c|}{}  	& \multicolumn{1}{c|}{}  			&  \ \  \ \ 0  \ \ 	 \ \ 			&  \\ \cline{2-4}
	Alice 						& \multicolumn{1}{c|}{ \ \ 0 \ \ } 	& \multicolumn{1}{c|}{1} 	& 1 	&  \\ \cline{2-4}
	& \multicolumn{1}{c|}{}  	& \multicolumn{1}{c|}{}  			& 0 						&  \\ \cline{1-4}
	\end{tabular}	
	\caption{Two possible winning instances for players Alice and Bob, if they are given $a=2$ and $b=3$, respectively. 
			They will win if Alice can fill the second row as $\{0,0,0\}$ and Bob the third column as  $\{0,0,1\}$, or $\{0,1,1\}$ and $\{0,1,0\}$ corresponding to the final matrices shown on left (i) and right (ii), respectively.}
	\label{table:ExampleWin}
\end{table}

 The shortcoming of utilizing classical resources in playing the MSG is that no matter what strategy they choose, the players can win against the referee only in eight cases out of nine, resulting in the average winning probability $8/9$. 
However, this winning probability could theoretically achieve unity if they could have shared a four-qubit entangled state given in Eq.\ref{eq:SuccessState}, and applied an appropriate quantum strategy  \cite{Brassard2005FoP}.

\begin{equation}
|\phi\rangle = {1 \over 2} ( |0011\rangle + |1100\rangle - |0110\rangle - |1001\rangle ),
\label{eq:SuccessState}
\end{equation}

\noindent where Alice holds the first two qubits and Bob holds the third and fourth qubits.
This four qubit state is actually the composition of two EPR  (Einstein-Podolsky-Rosen) pairs in the form ${1 \over \sqrt{2} }( |01\rangle - |10\rangle ) \otimes {1 \over \sqrt{2} }( |01\rangle - |10\rangle )$, each shared by Alice and Bob, such that Alice (Bob) possesses the first and third (second and fourth) qubits. 
The strategy they determine before the game starts is as follows. According to the row (column) number given by the referee, Alice (Bob) applies one of the three two-qubit operations $A_a$ ($B_b$), where $a, b \in \{1, 2, 3\}$, given in Eq.\ref{eq:matA} (Eq.\ref{eq:matB}). That is, following the above example, Alice applies $A_2$, and Bob applies $B_3$.

\begin{eqnarray}
A_1 ={1 \over \sqrt{2}}
\left(
\begin{array}{cccc}
i &  \ \ 0 & 0 & 1 \\
0 & -i & 1 & 0 \\
0 &  \ \ i & 1 & 0 \\
1 &  \ \ 0 & 0 & i \\
\end{array}
\right)\!\!\! , \ \ \ \
A_2 = {1 \over 2} \left(
\begin{array}{cccc}
\ \  i & 1 & \ \  1 & \ \ i \\
-i & 1 &     -1 & \ \ i \\
\ \  i & 1 &     -1 & -i \\
-i & 1 & \ \  1 & -i \\
\end{array}
\right)\!\! , \ \ \ \
A_3 = {1 \over 2} \left(
\begin{array}{cccc}
-1 &    -1 &     -1 & \ \ 1 \\
\ \  1 & \ \ 1 &     -1 & \ \ 1 \\
\ \  1 &    -1 &  \ \ 1 & \ \ 1 \\
\ \  1 &    -1 &     -1 &    -1 \\
\end{array}
\right)\!\! ,
\label{eq:matA}
\end{eqnarray}
\begin{eqnarray}
B_1 = {1 \over 2}
\left(
\begin{array}{cccc}
\ \  i &    -i & \ \ 1 & \ \ 1 \\
-i &    -i & \ \ 1 &    -1 \\
\ \  1 & \ \ 1 &    -i & \ \ i \\
-i & \ \ i & \ \ 1 & \ \ 1 \\
\end{array}
\right)\!\! , \ \ \ \
B_2 = {1 \over 2} \left(
\begin{array}{cccc}
-1 & \ \  i & 1 & \ \ i \\
\ \  1 & \ \  i & 1 &    -i \\
\ \ 1 &     -i & 1 & \ \ i \\
-1 &     -i & 1 &    -i \\
\end{array}
\right) \!\! , \ \ \ \
B_3 = {1 \over \sqrt{2}} \left(
\begin{array}{cccc}
\ \  1 & 0 & \ \ 0 & 1 \\
-1 & 0 & \ \ 0 & 1 \\
\ \  0 & 1 & \ \ 1 & 0 \\
\ \  0 & 1 &    -1 & 0 \\
\end{array}
\right)\!\!.
\label{eq:matB}
\end{eqnarray}

\noindent Next, measuring their qubits, each obtains two classical bits and determine the third bit according to the parity conditions. 
Note that the measurement of each party do not provide a single result, but one of the possible results with some probability.
However, thanks to the entangled state and quantum strategy, in the ideal case where there is no noise and experimental imperfections, the results of Alice and Bob are found to be both satisfying the parity conditions and that the intersecting number is the same.
Following the same example ($a=2$, $b=3$), in addition to two instances given in Table \ref{table:ExampleWin}, the instances (iii) through (viii) given in Table \ref{table:ExampleWin2} could occur each with $1/8$ probability, summing up to unity.
For a more detailed example, let us take instance (iv), that after applying $A_2$ and $B_3$, measurement results yield two classical bits $\{0,1\}$ for Alice and $\{1,1\}$ for Bob.
To satisfy the parity conditions, Alice extends her two-bit string to $\{0,1,1\}$, and Bob to $\{1,1,1\}$.

\begin{table}[b!]
	\centering
	\begin{tabular}{|c|ccc|l}
		\cline{1-4}
		iii &                      	& Bob                    			&   						&  \\ \cline{1-4}
		& \multicolumn{1}{c|}{}  	& \multicolumn{1}{c|}{}  			&  \ \  \ \ 1  \ \ 	 \ \ 			&  \\ \cline{2-4}
		Alice 						& \multicolumn{1}{c|}{ \ \ 0 \ \ } 	& \multicolumn{1}{c|}{0} 	& 0 	&  \\ \cline{2-4}
		& \multicolumn{1}{c|}{}  	& \multicolumn{1}{c|}{}  			& 0 						&  \\ \cline{1-4}
	\end{tabular}\!\!\!\!, \ \ 
	\begin{tabular}{|c|ccc|l}
		\cline{1-4}
		iv &                       	& Bob                    			&   						&  \\ \cline{1-4}
		& \multicolumn{1}{c|}{}  	& \multicolumn{1}{c|}{}  			&  \ \  \ \ 1  \ \ 	 \ \ 			&  \\ \cline{2-4}
		Alice 						& \multicolumn{1}{c|}{ \ \ 0 \ \ } 	& \multicolumn{1}{c|}{1} 	& 1 	&  \\ \cline{2-4}
		& \multicolumn{1}{c|}{}  	& \multicolumn{1}{c|}{}  			& 1 						&  \\ \cline{1-4}
	\end{tabular}\!\!\!\!, \ \ 
\begin{tabular}{|c|ccc|l}
\cline{1-4}
v &                      	& Bob                    			&   						&  \\ \cline{1-4}
& \multicolumn{1}{c|}{}  	& \multicolumn{1}{c|}{}  			&  \ \  \ \ 0  \ \ 	 \ \ 			&  \\ \cline{2-4}
Alice 						& \multicolumn{1}{c|}{ \ \ 1 \ \ } 	& \multicolumn{1}{c|}{0} 	& 1 	&  \\ \cline{2-4}
& \multicolumn{1}{c|}{}  	& \multicolumn{1}{c|}{}  			& 1 						&  \\ \cline{1-4}
\end{tabular}\!\!\!\!, \\
\

\ \\
	\ \ \begin{tabular}{|c|ccc|l}
	\cline{1-4}
	vi &                       	& Bob                    			&   						&  \\ \cline{1-4}
	& \multicolumn{1}{c|}{}  	& \multicolumn{1}{c|}{}  			&  \ \  \ \ 1  \ \ 	 \ \ 			&  \\ \cline{2-4}
	Alice 						& \multicolumn{1}{c|}{ \ \ 1 \ \ } 	& \multicolumn{1}{c|}{0} 	& 1 	&  \\ \cline{2-4}
	& \multicolumn{1}{c|}{}  	& \multicolumn{1}{c|}{}  			& 1 						&  \\ \cline{1-4}
\end{tabular}\!\!\!\!, \ \ 
\begin{tabular}{|c|ccc|l}
	\cline{1-4}
	vii &                      	& Bob                    			&   						&  \\ \cline{1-4}
	& \multicolumn{1}{c|}{}  	& \multicolumn{1}{c|}{}  			&  \ \  \ \ 0  \ \ 	 \ \ 			&  \\ \cline{2-4}
	Alice 						& \multicolumn{1}{c|}{ \ \ 1 \ \ } 	& \multicolumn{1}{c|}{1} 	& 0 	&  \\ \cline{2-4}
	& \multicolumn{1}{c|}{}  	& \multicolumn{1}{c|}{}  			& 1 						&  \\ \cline{1-4}
\end{tabular}\!\!\!\!, \ \ 
\begin{tabular}{|c|ccc|l}
	\cline{1-4}
	viii &                     	& Bob                    			&   						&  \\ \cline{1-4}
	& \multicolumn{1}{c|}{}  	& \multicolumn{1}{c|}{}  			&  \ \  \ \ 1  \ \ 	 \ \ 			&  \\ \cline{2-4}
	Alice 						& \multicolumn{1}{c|}{ \ \ 1 \ \ } 	& \multicolumn{1}{c|}{1} 	& 0 	&  \\ \cline{2-4}
	& \multicolumn{1}{c|}{}  	& \multicolumn{1}{c|}{}  			& 0 						&  \\ \cline{1-4}
\end{tabular}
	\caption{Given $a=2$ for Alice and $b=3$ for Bob under ideal conditions, in addition to two possible winning instances given in Table \ref{table:ExampleWin}, any of these six winning instances can occur each with probability $1/8$ by applying $A_2$ and $B_3$ and performing the measurement.}
	\label{table:ExampleWin2}
\end{table}

 However, quantum systems are very fragile that any source of imperfections during the process might affect the performance of the task, and the MSG is of no exception. 
Gawron et al.'s work on the noise effects in the MSG \cite{Gawron2008IJQI} clearly showed that if qubits hold by Alice and Bob are subject to noise, that their four-qubit state is not the pure state in Eq.\ref{eq:SuccessState} but rather a mixed state, the average winning probability decreases and with increasing noise, the probability can drop well below the classical limit $8/9$.
This work was followed by others in various settings \cite{Ramzan2010QIP,Fialik2012QIP,Gawron2016ActaB,Pawela2013Plos1,Ozaydin2020LP}.
Hence, although quantum advantage is imminent in theory, it is of great interest to design a physical system to bring this advantage to life and investigate the conditions for surpassing the classically achievable limit.

\noindent In this work, addressing this problem, we propose  a few nanometer-sized silicon single quantum dots (SQDs) \cite{garoufalis2001high,wilcoxon1999optical,wolkin1999electronic} with bandgap small enough to allow spin-photon interaction based setup within the reach of current technology. Electron spins confined in quantum dots provide a promising basis for quantum computation with  potential for scaling and reasonably long coherence time \cite{Loss_PRA_1998, DiVincenzo_nature_2000, taylor2005fault,Veldhorst_2014,leon2020coherent}. In the basic proposal \cite{Loss_PRA_1998}, single spins form a logical basis with a single qubit operation via spin resonance. The silicon-based quantum dot has been studied intensively and attracted great interest thanks to its charge offset stability and compatibility with CMOS and quantum information technology \cite{zimmerman2001excellent,fujiwara2001manipulation, dutta2000electron, takahashi2002silicon, ono2005manipulation, bugu2020rf}. Hence, progressive approaches based on quantum dots have been proposed in various areas of quantum information such as preparing  multipartite entanglement via Pauli spin blockade in double quantum dot system \cite{bugu2020}, and coupling photonic and electronic qubits in microcavity systems \cite{Han_SciRep_2015, Li_QIP_2015}. What is more, coupling quantum dots to nanophotonic waveguide \cite{Uppu2020}, and optical microcavity \cite{najer2019gated} for quantum information processing have recently been experimentally demonstrated. By considering various physical imperfections, we first show that the MSG can be implemented in a quantum system outperforming the classical resources under realistic conditions. 
Next, thanks to our physical analysis, we design a new version of the game, that having information on the physical realization and strategy of the players, in order to decrease their winning probability, referee can bias the game.

\begin{figure}[b!]
	\centerline{\includegraphics[width=0.8\columnwidth]{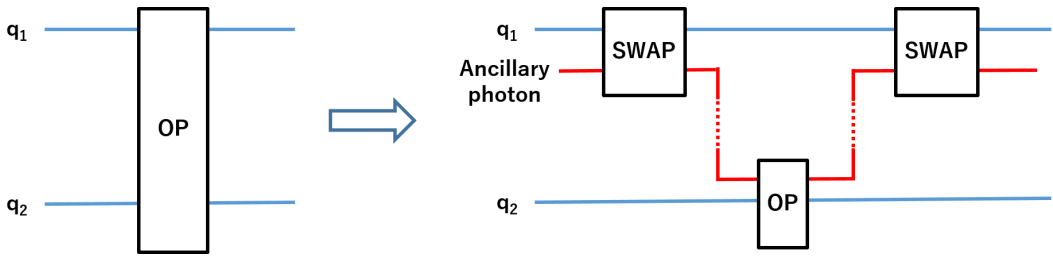}}
	\caption{Extending each two-qubit operation on two spin qubits denoted as $q_1$ and $q_2$, to a three-qubit operation via an ancillary photon, so that any two-qubit operation could be realized on spatially separated spin qubits.}\label{fig:Extension}
\end{figure}


\section*{Results}
Our setup is based on quantum computation with quantum dots coupled to spatially separated optical cavities.
In our setup, each spin of a quantum dot constituting each logical qubit of Alice and Bob is coupled to the optical field of the cavity. 
Introducing ancillary photons, quantum operations on two distant qubits of each player are realized through photon-spin interactions. 
That is, as illustrated in Fig.\ref{fig:Extension} each two-qubit operation on logical qubits is extended to an equivalent three-qubit operation which is realized by only single-qubit operations on photons or spins, and two-qubit operations on photon-spin pairs.
As already considered in many works \cite{Cheng2013JOSAB} and explained in Methods section, our configuration realizes a controlled-phase $CP(\pi-\theta)$ gate between spin and photon, which reduces to a controlled-Z ($CZ$) gate in the ideal condition for $\theta=0$, with $CZ=CP(\pi)$.
Hence, before considering physical imperfections and taking into account the effect of finite $\theta$ on winning probability, we first decompose each two-qubit unitary operation in terms of single-qubit gates (detailed in Methods section) and CZ gates, and then extend the decomposed two-qubit circuits to three-qubit circuits.

\noindent As controlled-Not (CNOT) gates with single qubit gates constitute a universal set, any unitary operation can be decomposed in terms of these gates \cite{nielsen2001quantum, iten2019introduction}.
What is more, a CNOT is equivalent to a ${\rm CZ}$ up to two Hadamard ($Had$) gates applied to the target qubit before and after the ${\rm CZ}$ gate, i.e. 
$CNOT^{1,2} \equiv  Had^2.CZ.Had^2$, where the superscript of single qubit gates represent which qubit it is applied to, and we use ${\rm CNOT}^{1,2}$ for the first qubit being the control and second qubit being the target qubit, and ${\rm CNOT}^{2,1}$ for the opposite case.
For each operation $A_a$ (and $B_b$), we find the decomposition $A^d_a$ in terms of ${\rm CNOT}$ and single qubit gates, and then the extension to three-qubit circuit $A^e_a$, and finally the circuit  $A^{CZ}_a$ consisting of only ${\rm CZ}$ and single qubit gates. We are now ready to present the decompositions we find for the two-qubit operations given in Eqs.\ref{eq:matA} and \ref{eq:matB}, as

\begin{figure}[t!]
		\includegraphics[width=18.0cm]{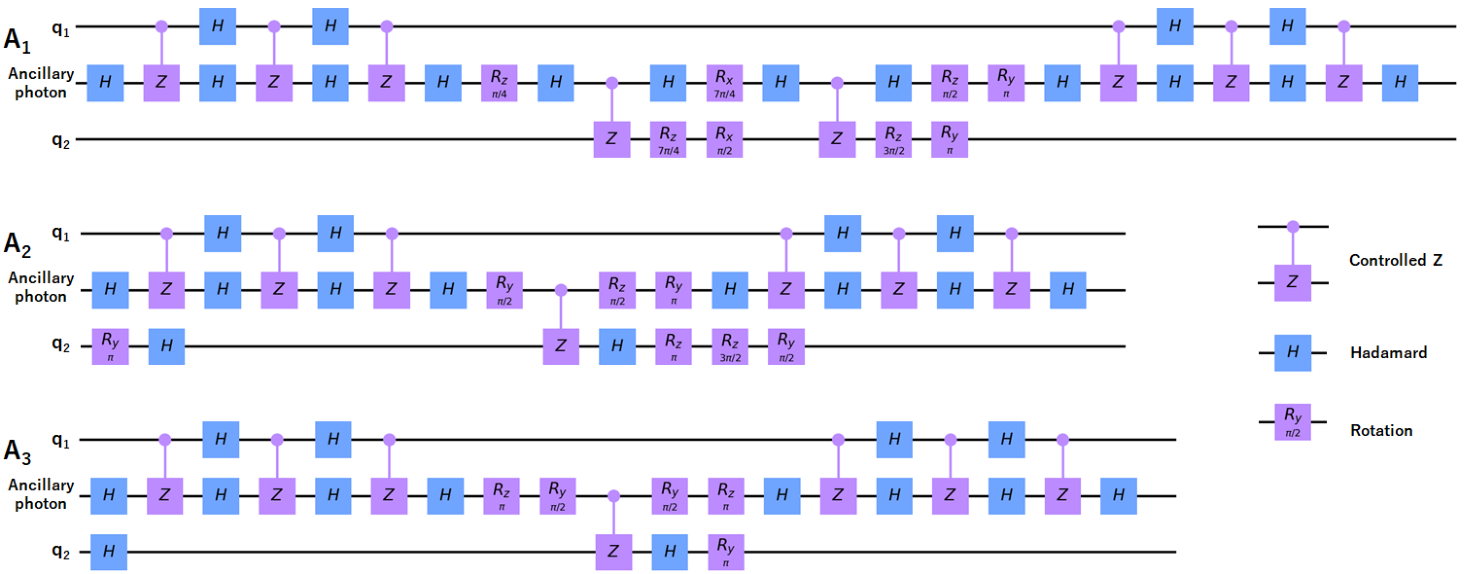} \\	
		\ \\
		\ \\
					
		\includegraphics[width=1.0\columnwidth]{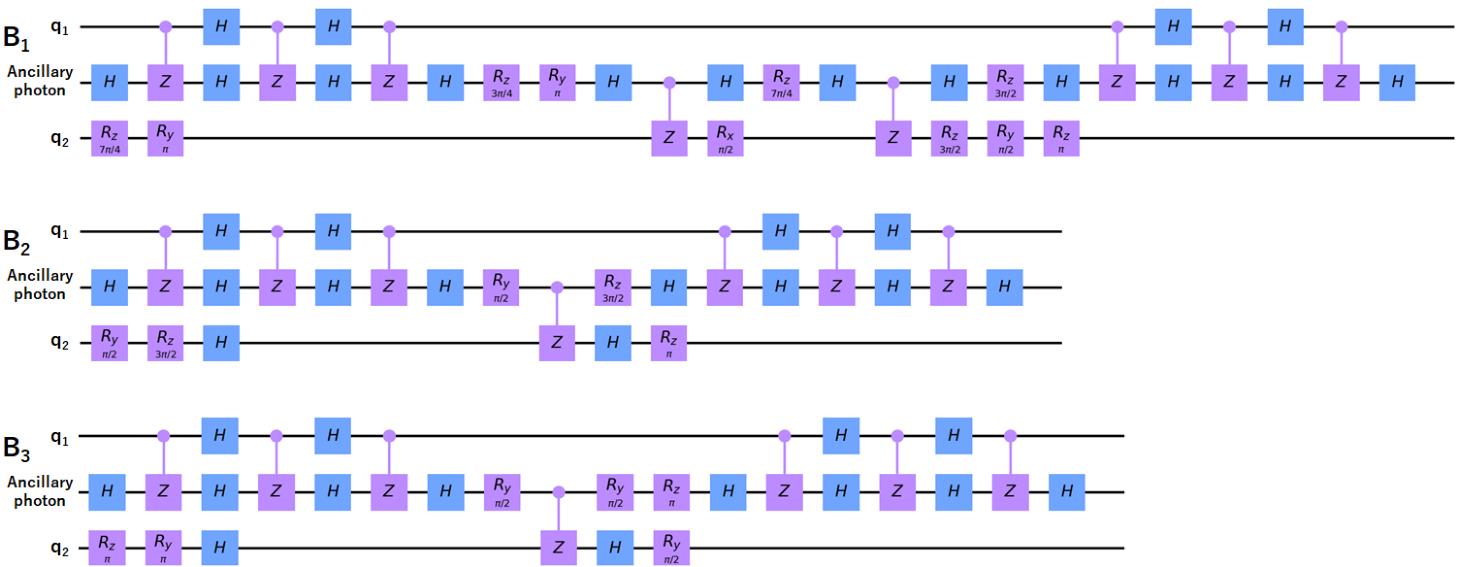} \\
	\caption{
		Circuit diagrams for realizing the operations of players. Blue $H$ gate represent a Hadamard gate, purple gates represents Controlled-Z (CZ) gate or $R_x$, $R_y$, and $R_z$ gates which are rotation around x, y, and z axis, respectively.
		Referee gives the row number $a$ to Alice and column number $b$ to Bob. 
		Alice applies $A_a$ and Bob applies $B_b$, each to his/her two spin qubits ($q_1$ and $q_2$) in distant optical cavities through an ancillary photon.
		As the photon passes through the cavity, the interaction realizes a CZ operation between the photon and spin in the ideal conditions. We used IBM qiskit \cite{cross2018ibm} to draw our decomposed circuit diagrams.
	}\label{fig:circuitAB}
\end{figure}

\begin{figure}[t!]
	\centerline{\includegraphics[width=0.8\columnwidth]{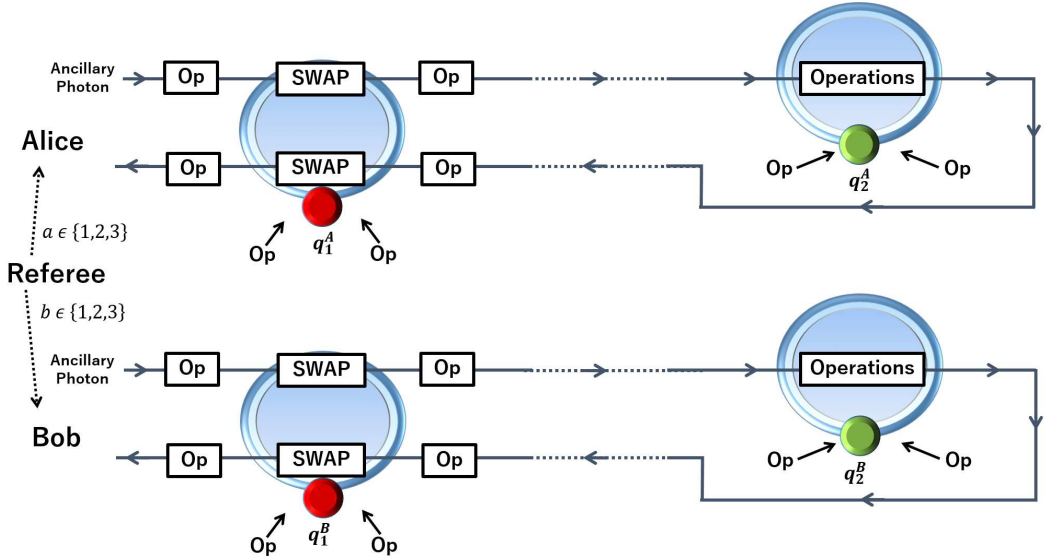}}
	\caption{Proposed experimental setup for playing the game. Referee gives the number of row $a$ (column $b$) to Alice (Bob) to fill with binary entries. 
		The initial four-qubit state given in Eq.\ref{eq:SuccessState} is a composition of two EPR pairs (one illustrated with red and the other with green circles) shared by Alice and Bob. 
	Each quantum dot (red or green circles) coupled to an optical cavity (blue toroids) constitutes one logical qubit.
	Following the extension strategy in Fig.\ref{fig:Extension}, each two-qubit operation (given in Eqs.\ref{eq:matA} and \ref{eq:matB}) on the logical qubits (distant quantum dots) is realized via an ancillary photon traveling between the optical cavities as: Following a SWAP operation between the photon and the spin coupled to the first cavity, the photon is sent to the second cavity to realize the desired operation. ``Operations'' represent the overall operations as decomposed in Eqs.\ref{eq:matA} red and \ref{eq:matB}, each containing single qubit operations, and one or two CZ operations. Each CZ is realized through the interaction between ancillary photon and second spin qubit, $q^A_2$ (or $q^B_2$).
	Each ``Op'' represents either an identity operator, or a set of single qubit operations on photonic or spin qubit. 
	After the ``Operations'', the photon is sent back to the first cavity for swapping back the quantum state with the spin qubit. Two-spin qubits of each party are now ready to be measured for obtaining the binary entries.}\label{fig:Setup}
\end{figure}

\begin{eqnarray}
A^d_1 &=&  e^{ - {7\pi \over 8} i}  R_z^1({\pi / 4}).CNOT^{2,1}. R_z^2({7\pi / 4})  . R_x^2({\pi / 2}). R_z^1({7\pi / 4}). CNOT^{2,1}. R_z^1({\pi / 2}). R_y^1({\pi }).  R_z^2({3\pi / 2}). R_y^2({\pi}), \\
A^d_2 &=& R_y^1({\pi / 2}) . R_z^2({\pi}) . CNOT^{1,2} . R_z^1({\pi / 2}). R_y^1({\pi}) . R_z^2({\pi}). R_y^2({\pi / 2}) . R_z^2({ 3 \pi / 2}), \\
A^d_3 &=& R_z^1(\pi) . R_y^1(\pi / 2) . CNOT^{1,2} . R_y^1(\pi /2). R_z^1(\pi) . R_y^2(\pi ), \\
B^d_1 &=& e^{ {7\pi \over 8} i} R_x^2(3 \pi / 2) . R_y^2(3 \pi / 4) . CNOT^{1,2} . R_z^1(\pi / 4) . R_y^1(3 \pi / 2). R_y^2(3 \pi / 2). CNOT^{1,2} . R_z^1(2 \pi ) . R_z^2(3 \pi / 2) . R_y^2(\pi), \\
B^d_2 &=& e^{i \pi} R_y^1( \pi / 2). R_y^2( \pi / 2). R_z^2(3 \pi / 2). CNOT^{1,2} . R_z^1(3 \pi / 2). R_z^2( \pi),\\
B^d_3 &=& R_y^1( \pi / 2). R_z^2( \pi). R_y^2( \pi). CNOT^{1,2}. R_y^1( \pi / 2). R_z^1( \pi ) . R_y^2( \pi / 2).
\end{eqnarray}

\noindent Unlike other four operations, requiring not one but two CNOT gates in the decomposition, $A_1$ and $B_1$ are going to play a significant role in the physical realization of the task, and give rise to a new version of the game.

For extending the decomposed two-qubit (spin-spin) circuits to three-qubit (spin-photon-spin) circuits as illustrated in Fig.\ref{fig:Extension}, we will make use of two-qubit SWAP gates, which can be realized as 
$SWAP \equiv CNOT^{1,2}.CNOT^{2,1}.CNOT^{1,2}$.
Our strategy for realizing the interaction between two spins via a three-qubit operation using only two-qubit gates is as follows.
For each player, the ancillary photon is sent to the first cavity to interact several times. 
Before and after each interaction which realizes a CZ gate between the photon and the spin, Hadamard gates are applied to both qubits appropriately, so that three CNOT gates equivalent to a SWAP gate are realized. That is, quantum states of the first spin and ancillary photon are swapped.
The photon is then sent to the other cavity containing the second spin.
Through interactions realizing CZ gates, and single qubit operations on spin and photon, the actual operation is realized.
Finally, the photon is sent back to the first cavity to swap back the quantum state with the spin.
The overall operation is equivalent to the corresponding two-qubit operation of the player.
We illustrate the circuit diagram for each overall operation in Fig.\ref{fig:circuitAB}, red and the corresponding experimental setup in Fig.\ref{fig:Setup}. 

\noindent We start with the initial state (in Eq.\ref{eq:SuccessState}) and two ancillary photons each in $|0\rangle$ state in the physical order of qubits as $Alice^1, Ancilla^{A}, Alice^2$ and $Bob^1,Ancilla^B,Bob^2$, which can be written as 

\begin{equation}
|\Psi\rangle = (SWAP \otimes id \otimes id \otimes SWAP) .(|0\rangle^{A} \otimes |\phi\rangle \otimes |0\rangle^{B}),
\label{eq:ExtendedSuccessState}
\end{equation}

\noindent where $id$ is the single qubit identity operator. 
Note first that this writing is only for the sake of clarity to explain the physical order of the qubits, hence the SWAP operations in Eq.\ref{eq:ExtendedSuccessState} are not taken into account in the physical realization.
Note also that for simplicity in tracing out operations during calculations, we start by swapping the ancillary photon with the first photon of Alice, while we swap the ancillary photon with the second photon of Bob.
With $a,b \in \{1,2,3\}$, extended three-qubit operations $A^e_a$ and $B^e_b$ are defined as

\begin{eqnarray}
A^e_a &=& (SWAP \otimes id) . (id \otimes A^d_a). (SWAP \otimes id), \\ 
B^e_b &=& (id \otimes SWAP) . (B^d_b \otimes id). (id \otimes SWAP).
\end{eqnarray}

\noindent Upon receiving the number $a$ ($b$) from the referee, Alice (Bob) applies the operation $A^e_a$ ($B^e_b$).
Next step is to trace out the ancillary qubits, and finally perform the measurements.
Under ideal conditions, as the ancillary qubits are back in their initial $|0\rangle$ states separable from the logical qubits, tracing them out will not disturb them. 
Note that as these measurements are not Bell measurements, they can be performed on distant spin qubits separately.

\subsection*{Physical Imperfections}

\begin{figure}[t!]
	\centerline{\includegraphics[width=0.8\columnwidth]{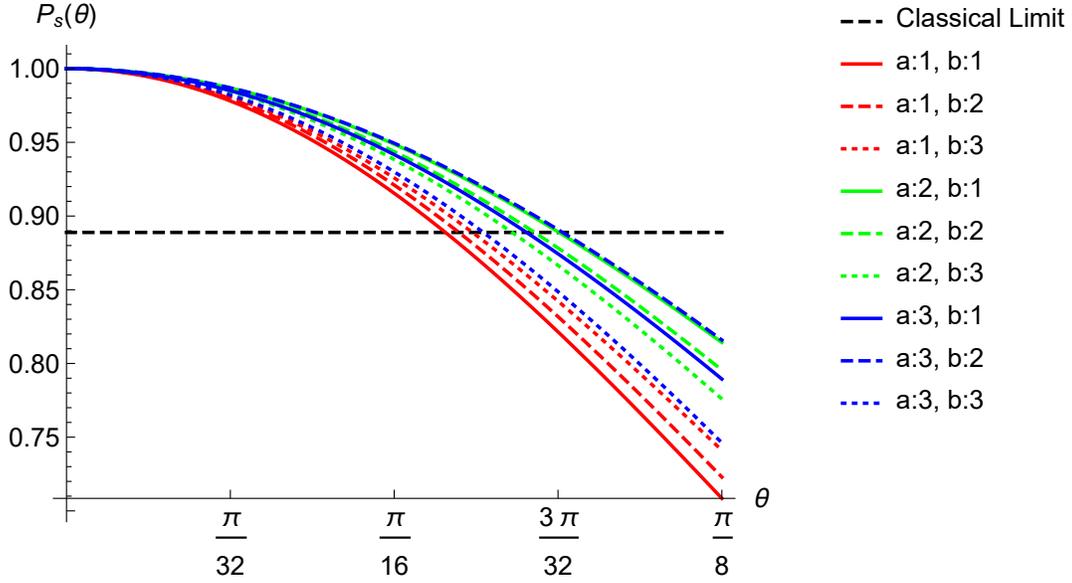}}
	\caption{Success probability $P_s$ as a function of $\theta$ for each pair of numbers $\{a, b\}$ referee can give Alice and Bob for filling the row and column, respectively, with binary entries. 
	Here, $\theta$ represents the imperfection of the interaction between the logical qubit and the ancillary qubit, i.e. realizing not the desired $CZ \equiv CP(\pi)$ but $CP(\pi-\theta)$ operation. 
	Operations $A_1$ and $B_1$ (corresponding to $a=1$ and $b=1$, respectively) are more complex than others that they contain more controlled operations, i.e. $CP(\pi-\theta)$. Hence, success probability of players decreases faster for $a=1$ or $b=1$, and the fastest for $a=b=1$.}\label{fig:mainresult1}
\end{figure}

Neglecting the imperfections such as the decoherence or absorption of photons between distant cavities, major physical imperfections we take into account in this analysis are due to Q-factor of the chosen optical cavity, coherence of qubits and the coupling, which all contribute to the imperfection in the desired operation between spin qubits and photon qubits. 
In general, according to the technology used, the realized operation might deviate from the ideal $CZ \equiv CP(\pi)$ to $CP(\pi-\theta)$.
Following our decomposition and extension, it is straightforward to take this effect into account by simply replacing each CZ in the circuits with $CP(\pi-\theta)$.
This time, final states of the ancillary qubits will not be $|0\rangle$, and measurement result on each qubit pair will not yield $1/8$, but rather a function of $\theta$.
We plot the success probability $P_s$ for each $\{a,b\}$ in Fig.\ref{fig:mainresult1}.
As we expected, $a=1$ and $b=1$ is the worst case for players. 
On the contrary to the other cases, the decompositions of both $A_1$ and $B_1$ contain not one but two CNOT gates, they require two imperfect CZ gates in the realistic settings, which lead to a potential new version of MSG.

\section*{Discussion}

Following the usual scenarios in quantum games, we assumed that Alice and Bob initially shared the ideal state given in Eq.(\ref{eq:SuccessState}), and did not take into account the imperfections in preparing the state, which could slightly decrease the overall success probabilities. 
However, as the initial state consists of two Bell states, its preparation is straightforward \cite{Munro2008PRB1,Munro2008PRB2}. 
On the other hand, Alice and Bob could choose to prepare the initial state not based on four spin qubits, but two being the photonic qubits. Hence, the first SWAP gate (i.e. the imperfect CZ gates to realize it) could be removed, this time increasing the overall success probability. 

For realizing the interaction between the incident photon and the spin qubit in not only in quantum dots but in also nitrogen-vacancy centers in diamond, and also atomic qubits, various optical microcavities are considered. 
Achieving ultra-high quality factor, microtoroid resonators with whispering gallery modes are promising \cite{wei2015hybrid,chen2011entangling,cheng2013quantum, wei2013compact}. 
Single-sided or double-sided cavities even with small Q-factors \cite{an2009quantum, li2019high} are also shown to be candidates for realizing atom-photon or spin-photon interactions with high success rate, enabling myriad quantum information processing tasks from entanglement generation to quantum teleportation (see Refs \cite{Munro2008PRB1,Munro2008PRB2, duan2004scalable, 
	heo2017implementation} and references therein).

Note that, our particular setup herein is robust because due to Eq.\ref{eq:refcoeff1}, the parameter $\theta$ can take only two values, $\pi$, or $0$, the latter being either for realizing the desired operation according to the conditions (as explained in the Methods section), or deviating from $\pi$ only in the extremely imperfection conditions such as $g \ll 5 \sqrt{\kappa \gamma}$ (where $g$ is the coupling strength of the cavity to the quantum dot, $\kappa$ is the cavity decay rate and $\gamma$ is the quantum dot spin decay rate), which is not anticipated with high-Q resonators.
However, our analysis on the affect of physical imperfections on the success probability is more general for realization in any technology where the interaction yields the operation $CP(\pi-\theta)$ with a finite $\theta$.

Our analysis showed that, if some of the possible cases for realizing a task requires more complicated operations, new versions of the task could arise.
Suppose Alice and Bob are playing MSG against a referee, with the present experimental setup following our decomposition. 
Then the question lying in the heart of game theory is, whether the referee has information on their setup and strategy. 
If so, in order to decrease the winning probability of the players (that is increasing the own winning probability), instead of each round with an evenly distributed random numbers $a$ and $b$ for the row and column, respectively, the referee can tend to choose always $a=1$ or $b=1$, and even $a=b=1$.

In summary, taking into account possible physical imperfections, we proposed a physical setup for playing MSG feasible with the current technology. 
We found the limits of imperfections for surpassing the classical winning probability.
We also showed that, depending on the partial information, the referee can bias the game to increase his/her winning probability, which gives rise to a new version of the Magic Square game.

\section*{Methods}

\begin{figure}[t]
	\centerline{\includegraphics[width=0.2\columnwidth]{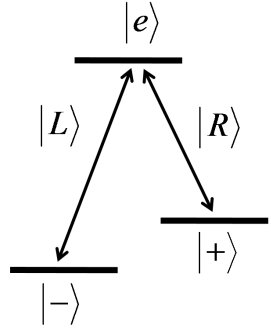}}
	\caption{$\Lambda$ type optical transitions possible in a quantum dot. The transitions $|-\rangle  \leftrightarrow |e\rangle$ and $|+\rangle \leftrightarrow |e\rangle$  are associated with the left and right polarization of the photon, denoted as $|L\rangle$ and $|R\rangle$ respectively.}\label{fig:excitation}
\end{figure}

A quantum dot coupled to an optical cavity can be coupled to the cavity mode, and the interaction between the cavity and the quantum dot spin is governed by Jaynes-Cummings model with the Hamiltonian

\begin{equation}
	H = \sum_{j=R,L} [ {\omega_{j0} \over 2} \sigma_{jz} + \omega_{jC} a^{\dagger}_j a_j + i g_j(a_j \sigma_{j+} - a^{\dagger}_j \sigma_{j-})] + H_R,
\end{equation}

\noindent where $a^{\dagger}$ and $a$ are the creation and annihilation operators of the cavity field, respectively. $R$ and $L$ denote the circular polarizations of the photon, associated with the optical transitions in the quantum dot (see Fig.~\ref{fig:excitation}) and index $j$ runs for $R$ and $L$.
$\omega_0$ and $\omega_C$ are the transition frequencies of the electronic energy levels and the frequency of the cavity field, $\sigma_+$, $\sigma_-$ and $\sigma_z$ are the raising, lowering and inversion operators of the quantum dot spin between the two corresponding levels, respectively.
The Hamiltonian for the field and atomic reservoirs are denoted by $H_R$, and we take $\hslash =1$.
Applying a magnetic field, non-zero spin level splitting can be achieved, so that R and L polarized photons receive different phase shifts upon the interaction with the quantum dot-cavity system \cite{Munro2008PRB1}, as explained below.

When an incident photon with frequency $\omega_p$ is introduced to the cavity, the Langevin equations for $a$ and $\sigma_-$ can be obtained for the low temperature reservoir and neglecting the vacuum input field, as

\begin{equation}
	{  \text{d} a_j \over \text{d}t} = [ i (\omega_p - \omega_C) - { \kappa \over 2} ] a_j(t) - g\sigma_{j-}(t) - \sqrt{\kappa} a_{j, in} (t),
\end{equation}

\begin{equation}
	{  \text{d} \sigma_{j-} \over \text{d}t} = [ i (\omega_p - \omega_0) - { \gamma \over 2} ] \sigma_{j-}(t) - g\sigma_{j,z}(t) a_{j} (t),
\end{equation}

\noindent where $g$ is the coupling strength of the cavity to the quantum dot, $\kappa$ is the cavity decay rate and $\gamma$ is the quantum dot spin decay rate.
Assuming weak assumption limit $\langle \sigma_z \rangle = -1$ and adiabatically eliminating the cavity mode, the reflection coefficient for the input photon pulse is found as ~\cite{Milburn94,Munro2008PRB2}

\begin{equation}\label{eq:refcoeff}
	r(\omega_p) = {  [ i (\omega_C - \omega_p) - {\kappa \over 2}  ] [ i (\omega_0 - \omega_p) + {\gamma \over 2} ] + g^2
		\over
		[ i (\omega_C - \omega_p) + {\kappa \over 2}  ] [ i (\omega_0 - \omega_p) + {\gamma \over 2} ] + g^2 }.
\end{equation}

\noindent If the quantum dot is uncoupled from the cavity, the reflection coefficient for the input photon becomes

\begin{equation}\label{eq:refcoeff0}
	r_0(\omega_p) = {   i (\omega_C - \omega_p) - {\kappa \over 2}
		\over
		i (\omega_C - \omega_p) + {\kappa \over 2}
	}.
\end{equation}

\noindent The reflection coefficients can be obtained for the resonant condition $\omega_p=\omega_0=\omega_C$ as

\begin{equation}\label{eq:refcoeff1}
	r(\omega_p) = {    - \kappa \gamma + 4 g^2
		\over
		\kappa \gamma + 4 g^2
	}, \ \ \text{and} \ \  r_0(\omega_p) = -1.
\end{equation}

Due to the spin-dependent optical transition rules \cite{Cheng2013JOSAB} as simply illustrated in Fig.~\ref{fig:excitation}, and optical Faraday rotation, an $|R\rangle$ polarized incident photon receives a phase shift $e^{i \phi_0}$ because, due to large level splitting, the spin state of the quantum dot is decoupled from the incident pulse \cite{Munro2008PRB1}. However, if the incident photon is $|L\rangle$ polarized, it will receive a phase shift $e^{i \phi}$ ($e^{i \phi_0}$) depending on the spin state of the quantum dot $|-\rangle$ $(|+\rangle)$, where $\phi$ and $\phi_0$ are the arguments of $r(\omega_p)$ and $r_0(\omega_p)$, respectively. For the resonant condition, and $g > 5 \sqrt{\kappa \gamma}$, one approximately finds $\phi=0$ and $\phi_0=\pi$. Placing a $\pi$ phase shifter to the photon reflection 
path, a controlled-Z gate between the electronic spin of the quantum dot and the incident photon is realized as $|R\rangle|+\rangle \rightarrow |R\rangle|+\rangle$, $|R\rangle|-\rangle \rightarrow |R\rangle|-\rangle$, $|L\rangle|+\rangle \rightarrow |L\rangle|+\rangle$, $|L\rangle|-\rangle \rightarrow -|L\rangle|-\rangle$.
The implementations of single qubit operations on spins and incident photons can be realized effectively and with high fidelity via electric pulses \cite{yoneda2018quantum} and half wave plates \cite{bartkowiak2010linear}, respectively.
One- and two-qubit operations we use in this work are

\begin{equation}
\text{CNOT} =
\left(
\begin{array}{cccc}
1 &    0 &  0 &  0 \\
0 &    1 &  0 &  0 \\
0 &    0 &  0 &  1 \\
0 &    0 &  1 &  0 \\
\end{array}
\right), \ \ \ 
CP( \theta ) =
\left(
\begin{array}{cccc}
1 &    0 &  0 &  0 \\
0 &    1 &  0 &  0 \\
0 &    0 &  1 &  0 \\
0 &    0 &  0 & \exp({i \pi \theta })\\
\end{array}
\right),
\end{equation}
\begin{equation}
R_x( \theta ) =
\left(
\begin{array}{cc}
\cos {\theta \over 2}  & i \sin{\theta \over 2}  \\
i \sin{\theta \over 2} & \cos {\theta \over 2}  \\
\end{array}
\right),\ \ \ 
R_y( \theta ) =
\left(
\begin{array}{cc}
\cos {\theta \over 2}  & \sin{\theta \over 2}  \\
- \sin{\theta \over 2} & \cos {\theta \over 2}  \\
\end{array}
\right),\ \ \ 
R_y( \theta ) =
\left(
\begin{array}{cc}
\exp({i \theta \over 2})  & 0  \\
0 & \exp({ - i \theta \over 2})  \\
\end{array}
\right).
\label{eq:gates2}
\end{equation}

\bibliography{sample}

\section*{Acknowledgements}

SB thanks to Roger Colbeck and Raban Iten, and FO thanks to Bilen Basarir for fruitful discussions. SB acknowledges Japanese Government MEXT scholarship. FO acknowledges the Personal Research Fund of Tokyo International University. This work was partially supported by JSPS KAKENHI Grant Number: JP18K18996 and JP20H00237, JST CREST (JPMJ CR1675) and MEXT Quantum Leap Flagship Program (MEXT Q-LEAP) Grant Number JPMXS0118069228.

\section*{Author contributions statement}

S.B. designed the scheme and carried out the theoretical analysis under the guidance of F.O. and T.K., S.B., F.O. and T.K. reviewed the manuscript and contributed to the interpretation of the work and the writing of the manuscript.

\section*{Competing interests}

The authors declare no competing interests.

\end{document}